\documentstyle[12pt,epsfig]{article}
\date{July 24, 1998}
\title{
\vspace{-10mm}
\rightline{\small  UL--NTZ 14/98}
Production of $e^+e^-$ pairs  to all orders in $Z\alpha$ 
for collisions of high-energy muons with heavy nuclei 
} 
\author{
D.~Ivanov$^{1,2}$\thanks{Email address: Dmitri.Ivanov@itp.uni-leipzig.de} , 
E.A.~Kuraev$^3$\thanks{Email address: kuraev@thsun1.jinr.dubna.su} ,
A.~Schiller$^1$\thanks{Email address: schiller@tph204.physik.uni-leipzig.de} , 
and
V.G.~Serbo$^{2,4}$\thanks{Email address: serbo@math.nsc.ru} 
\\
{\it $^1$Institut f\"ur Theoretische Physik and NTZ,
Universit\"at Leipzig,}\\
{\it D-04109 Leipzig, Germany} \\
{\it $^2$Novosibirsk State University, Novosibirsk, 630090
Russia} 
\\
{\it $^3$Joint Institute of Nuclear Research, Dubna, 141980
 Russia}\\
{\it $^4$INFN, Via Celoria 16, I-20133 Milano, Italy}
}

\begin{document}

\maketitle

\begin{abstract}
{  
The $e^+e^-$ pair production in 
collisions of muon with atoms is considered to all orders
in the parameter $Z\alpha$. We obtain energy distribution of $e^+$ and $e^-$
as well as energy loss of muon passing through matter with heavy atoms.
The found corrections to the Born contribution do not depend
on the details of the target properties except of a simple factor.
For the considered example of Pb target the 
muon energy loss corrections 
vary from $-65$ \% to $-10$ \% depending on the pair energy.
}
\end{abstract}
 
\section{Introduction}

The production of $e^+e^-$ pairs in collisions of high energy muons with
nuclei and atoms is important for a number of problems. In particular,
this process is dominant for energy losses of muons passing through matter.
An precise knowledge of these losses is necessary for the  
construction of detectors and $\mu^+\mu^-$ colliders and an 
estimation of shielding at high energy colliders.  

In Born approximation various cross sections for the process under discussion
($A$ denotes an atom or a nucleus with charge number $Z$) 
\begin{equation}
\mu A \to \mu A \ e^+ e^-
\label{proc}
\end{equation}
have been calculated in Refs. \cite{LL}-\cite{N}.
A recent short review on the muon
energy loss at high energy can be found in Sect. 23.9 of \cite{PDG}.
Some useful approximate formulae and figures are given in \cite{G}.

 In all mentioned  papers effects of high order
corrections in the parameter 
\begin{equation}
\nu =Z\alpha  \approx \frac{Z}{137}
\label{1}
\end{equation}
have not been taken into account. However, this parameter  is of the
order of 1 for heavy nuclei ($\nu=0.6$ for Pb) and, therefore, the
whole series in $\nu$ has to be summed to achieve an exact result for process
(\ref{proc}).

\begin{figure}[!htb]
\begin{center}
\epsfig{file=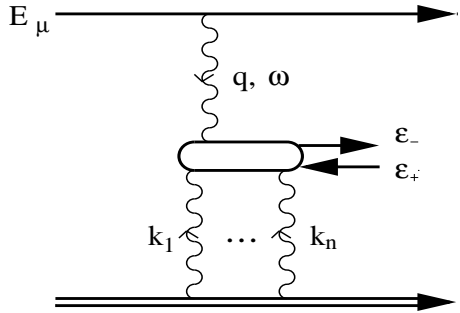,width=40mm,height=60mm,angle=270}
\caption{Amplitude $M_n$ with $n$ exchanged photons for reaction
$\mu A \to \mu A \ e^+ e^-$}
\end{center}
\end{figure}
Let $M_n$ be the amplitude of the discussed process with $n$ exchanged photons
(Fig.~1). 
We present the cross section in the form
\begin{eqnarray}
d \sigma&=& d \sigma_{\mathrm{Born}} + d \sigma_{\mathrm{Coul}}\,, 
\nonumber \\
 d \sigma_{\mathrm{Born}} \propto
\vert M_{\mathrm{Born}}\vert^2 = \vert M_1\vert^2 \,, &&   
d \sigma_{\mathrm{Coul}}\propto \vert\sum_{n=1}^{\infty} M_n\vert^2 - 
\vert M_1\vert^2
\label{cross}
\end{eqnarray}
where $M_{\mathrm{Born}}=M_1$ denotes the Born amplitude.
We call the Coulomb correction (CC) the difference $d\sigma_{\mathrm{Coul}}$
between the exact result 
and the Born approximation.

Such kind of CC is well known in the photoproduction of $e^+e^-$ pairs on
atoms (see \cite{BM}, \S 32.2 of \cite{AB} and \S 98 of \cite{BLP}). 
In case of the total cross section the 
corrections are negative and decrease the Born contribution by
 about 10 \% for Pb.

In the present paper we calculate CC for reaction (\ref{proc}) 
 neglecting only pieces of the order of 
\begin{equation}
\frac{m_\mu^2}{E_\mu^2}\,, \ \ \ \frac{m_e}{\varepsilon_\pm} \,.
\label{acc}
\end{equation}
Therefore, our results are valid for ultrarelativistic leptons.
In (\ref{acc}) $m_e$ and $m_\mu$ are the lepton masses,  $\varepsilon_\pm$
and $ E_\mu$ denote the lepton energies.

The discussed process for $\varepsilon_\pm \ll E_\mu$ has a close relation
to  $A'A \to A'A e^+e^-$ where $A'$ is a fast nucleus with
relatively small charge $Z' \alpha \ll 1 $ and $A$ is a heavy atom
or nucleus with $Z\alpha \sim 1$. The latter process was considered in
Refs. \cite{NP} and \cite{BB} assuming that the lepton energies are much smaller
than the energy of the projectile nucleus $A'$.
In \cite{NP,BB} the same complicated method has been used as in Ref.~\cite{BM}
which basically uses
approximated relativistic wave functions of $e^+$ and $e^-$ in the Coulomb
field of the nucleus $A$.

Our approach is more simple and transparent. It is based on  cross sections
for the virtual process $\gamma^* A \to e^+e^- A$ (where $\gamma^*$ denotes
the virtual photon with 4-momentum squared $ q^2 < 0$) which has been
recently obtained in Ref. \cite{IM}  by a direct summation of the corresponding
Feynman diagrams.
For $\varepsilon_\pm \ll E_\mu$ our Eqs.~(\ref{dsigcouly})  and
(\ref{sigmacut}) coincide with  Eqs. (38) and (39) 
of \cite{NP}, respectively, while our Eq.~ (\ref{dsigcoul})  
coincides with the corresponding equation
of \cite{BB} only in the main logarithmic approximation.
(It should be noted, however, that the results of \cite{BB} for the discussed
process are also  presented in  logarithmic accuracy.)

The outline of our paper is as follows:
In Sect.~2 we obtain the 
energy distributions for electrons and positrons. The next Sect. is devoted to 
the muon energy loss.
In Sect.~4 we summarise our results and 
compare the obtained Coulomb corrections with the Born
contributions.
Additionally we discuss
CC for similar  processes with other charged projectiles ($e$, $\pi$, $p$
instead of $\mu$).

Our main notations are collected in Fig.~1:  
$q$ and  $\omega$ 
are the 4-momentum and energy of the virtual photon generated by the
projectile muon and $k_1,\dots, k_n$ are the 4-momenta of the photons
exchanged with the nucleus.
Besides we use 
\begin{equation}
x_\pm= \frac{\varepsilon_\pm}{\omega} \,, \ \ \ 
x_++x_-=1 \,, \ \ \ y= \frac{\omega}{E_\mu}
\,, \ \ \  Q^2=-q^2 > 0 \,.
\end{equation}
Throughout the paper the well known function
\begin{equation}
f(\nu) = \nu^2 \sum_{n=1}^{\infty} {1\over n(n^2+\nu^2)}  
\label{12}
\end{equation}
and the abbreviation
\begin{equation}
\sigma_0=\frac{4}{3 \pi} \frac{Z^2 \alpha^4}{m_e^2} 
\label{sigma0}
\end{equation}
are used.

\section{Energy distribution of $e^+$ and $e^-$}

It is well known \cite{PDG} that the cross section 
for process (\ref{proc}) as well as for  electroproduction
can be exactly written in terms of two structure
functions or two cross sections $\sigma^T(\omega,
Q^2)$ and $\sigma^S(\omega, Q^2)$ for the
virtual processes 
$\gamma_T^* A  \to e^+e^- \ A$
and $\gamma_S^*A \to e^+ e^- \ A$
(where $\gamma_T^*$ and $\gamma_S^*$ denote the transverse
and scalar/longitudinal photons with helicity $\lambda_T=\pm 1 $ and
$\lambda_S=0$, respectively): 
\begin{equation}
d\sigma= \sigma^T(\omega, Q^2)\,
dn_T(\omega, Q^2)
+ \sigma^S(\omega, Q^2)\,dn_S(\omega, Q^2) \,.
\label{39}
\end{equation}
Here the coefficients $dn_T$ and $dn_S$ are called the number of
transverse and scalar virtual photons (generated by the muon)
with energy $\omega$ and virtuality $Q^2$, respectively. The cross
sections $\sigma^T$ and $\sigma^S$ 
have been calculated recently in Ref.~\cite{IM}\footnote{In Eqs. (46), (49)
and (59) of \cite{IM} a factor $x_+ x_-$ is missing in the integrands
 of  quantities 
$\sigma_1^S$ and $\sigma_2^S$.}:
\begin{eqnarray}
&&d \sigma^T= d \sigma_1^T + 
d \sigma_2^T = \frac{4}{3} \frac{Z^2 \alpha^3}{m_e^2}
\left[ L - f(\nu)\right] \left[ \frac{ m_e^4}{ (m_e^2+ Q^2 x_+x_-)^2}
+ \frac{ 2 (x_+^2+x_-^2) m_e^2}{ m_e^2+ Q^2 x_+x_-}
\right] d x_+ 
\,, \nonumber \\
&&d \sigma^S= d \sigma_1^S + 
d \sigma_2^S = \frac{4}{3} \frac{Z^2 \alpha^3}{m_e^2}
\left[ L - f(\nu)\right] \; \frac{ 4 m_e^2 Q^2 x_+^2 x_-^2 }
{ (m_e^2+ Q^2 x_+x_-)^2} \; dx_+ 
\label{dsig12}
\end{eqnarray}
with
\begin{equation}
L= \ln \frac{2 \omega x_+ x_-}{m_e} - \frac{1}{2} \ln 
\frac{ m_e^2 + Q^2 x_+ x_-}{m_e^2} - \frac{1}{2}  
\label{bigL}
\end{equation}
and the function $f(\nu)$ is given in Eq.~(\ref{12}).
The cross sections $d \sigma_1^{T,S}\propto L$ correspond to 
the Born contributions
and $d \sigma_2^{T,S}\propto - f(\nu)$ to CC.
The accuracy of cross sections (\ref{dsig12}) is determined omitting only 
pieces of the order of
\begin{equation}
\frac{ m_e}{\omega}\,, \ \ \ \frac{Q}{\omega} \,.
\label{acc2}
\end{equation}
The  number of photons can be found in Sect.~6 and App.~D 
of review \cite{BGMS}
(with accuracy ${\mathcal O}(m_\mu^2/E_\mu^2)$, ${\mathcal O}(Q^2/\omega^2)$)
\begin{eqnarray}
 d n_T&=& \frac{\alpha}{\pi} \left( 1-y \right) 
\left[ \left( 1 -\frac{Q^2_{\min}}{Q^2}\right) D +  \lambda \; C \right]
\frac{d \omega}{\omega}
\frac{d Q^2}{Q^2}\,,
\nonumber \\
 d n_S&=& \frac{\alpha}{\pi} \left( 1-y \right) 
\left[ \left( 1 +\frac{\lambda}{2}\right) D -  \frac{\lambda}{2} \; C \right]
\frac{d \omega}{\omega}
\frac{d Q^2}{Q^2} 
\label{numbers}
\end{eqnarray}
where
\begin{equation}
\lambda= \frac{1}{2} \frac{ y^2}{1-y} \,, \ \ \ 
Q^2_{\min} = \frac{y^2}{1-y} m_\mu^2\,, \ \ \ y=\frac{\omega}{E_\mu} \,.
\label{number1}
\end{equation}
For the considered case of muon projectile $C=D=1$, other particles are discussed in 
Sect.~4.
Eqs.~(\ref{39})-(\ref{number1}) are the basis for our following calculations.

Integrating Eq. (\ref{39}) over $Q^2$ from $Q^2_{\min}$ to infinity (the upper 
limit can be set to infinity due to the fast convergence of the integral)
we obtain the known Born contribution and the new expression for CC:
\begin{equation}
d \sigma_{\mathrm{Coul}}= - \sigma_0 f(\nu) \; F(x,y) \frac{d \omega}{\omega} 
d x_+ 
\label{dsigcoul}
\end{equation} 
with
\begin{eqnarray}
&&F(x,y)= (1-y) \left\{ \Large[ (1+\lambda + \xi) \; a -1-\lambda \Large] 
\ln \left(1+ \frac{1}{\xi} \right) -a  + \frac{ 4-a-\lambda}{1+\xi} \right\} \,,
\label{Fxy}
\\
&&a= 2 ( 1+x_+^2+x_-^2) \,, \ \ \ 
\xi= \frac{m_\mu^2}{m_e^2} \frac{y^2}{1-y} x_+ x_- \,.
\nonumber
\end{eqnarray}
Since the integration variables can be transformed as follows
\begin{equation}
\frac{d \omega}{\omega} dx_+ = 
\frac{ d \varepsilon_+ d \varepsilon _-}{\omega^2}
\end{equation}
Eq.~(\ref{dsigcoul}) describes the energy distribution of $e^+$ and $e^-$ in CC.
In the limit $\xi \ll 1$ (or $y \ll m_e/m_\mu$)  
the function $F(x,y)$ 
is approximated by 
\begin{equation}
F(x,y)= (1+ 2 x_+^2 + 2 x_-^2) \ln \frac{1}{\xi} - 4 ( x_+^2+x_-^2) \,.
\end{equation}
At $\xi\gg 1$ we obtain
\begin{equation}
F(x,y)=\frac{1}{\xi} [ 1-y+y^2+ 2 (1-y-y^2) x_+ x_-] \,.
\end{equation}
It is easy to see that 
the main contribution to $\sigma_{\mathrm{Coul}}$ arises from the region
\begin{equation}
m_e^2 \ll \varepsilon_+\varepsilon_- \ll 
\left( m_e \frac{E_\mu}{m_\mu}\right)^2 \,.
\label{main}
\end{equation}

Strictly speaking, the cross sections (\ref{dsig12}) are valid for pair 
production processes on nuclei.
In the collisions of  virtual photons with atoms, 
an atomic screening effect has to be taken into account.
For high energy photons the screening
effect changes considerably the differential and total cross section as well 
as the energy loss for the Born contribution. 
The reason is that the region of small
transverse momenta $k_{1\perp}
\stackrel{<}{\scriptstyle{\sim}} 1/a \sim m_e \alpha Z^{1/3}$ 
($a$ denotes the atomic radius)
significantly contributes to the cross sections.
As a consequence, the function in the Born contribution equivalent 
to our $F(x,y)$ becomes very complicated and not universal for different atoms 
(see \cite{N}).
On the contrary, the region  mainly contributing to CC, is determined by the
condition $k_{1\perp},\dots,k_{n\perp}\sim m_e \ll 1/a$.
Therefore, the atomic screening effect is negligible in CC
and the function $F(x,y)$ as well as some other distributions are universal and do not 
depend on  atomic properties.

However, if one is interested in very high energy pairs  
effects of the nucleus form
factor have to be taken into account both in the Born contributions
as well as in the Coulomb corrections.
This happens in the case when the characteristic squared momentum transferred
to the nucleus $\sim m_e^2 + Q^2_{\min} x_+ x_-$ becomes comparable with
$(1/R_A)^2$ where $R_A$ is the radius of the nucleus. From this condition it
follows that the just mentioned universal behaviour is spoiled for $y > 0.5$
where this pair production is strongly suppressed.

\section{Muon energy loss}
The Coulomb correction to the 
spectrum of the muon energy loss can be obtained from Eq.~(\ref{dsigcoul})
after integrating over $x_+$:
\begin{equation}
d \sigma_{\mathrm{Coul}}= - \sigma_0 \; f(\nu)\; F(y) \frac{dy}{y}\,, \ \ \
F(y)=(1-y) F_1(z) + y^2 F_2(z)
\,, \ \ \
z=\frac{ m_\mu^2}{m_e^2} \frac{y^2}{1-y} 
\label{dsigcouly} 
\end{equation}
where
\begin{eqnarray}
 F_1(z)&=&\frac{44}{15 z} - \frac{16}{15} - \left(\frac{7}{3}+\frac{8 z}{15}
\right) \ln z +
\nonumber \\
&  +&\left(-\frac{44}{15 z}+\frac{4}{4+z}+\frac{38}{15}+\frac{16z}{15}\right)
\sqrt{1+\frac{4}{z}} \; \ln \left( \sqrt{1+\frac{z}{4}}+\sqrt{\frac{z}{4}}\right)
\,, \nonumber \\
F_2(z)&=&-\frac{4}{3 z} - \frac{7}{6} \ln z + 
\nonumber \\
&+&
\left(-\frac{2}{3z} + \frac{8}{z(4+z)} +\frac{7}{3} \right)
\sqrt{1+\frac{4}{z}} \; \ln \left( \sqrt{1+\frac{z}{4}}+\sqrt{\frac{z}{4}}\right)
\,.
\label{F1F2}
\end{eqnarray}
The function $F(y)$ is presented in Fig.~2.
\begin{figure}[!htb]
\begin{center}
\epsfig{file=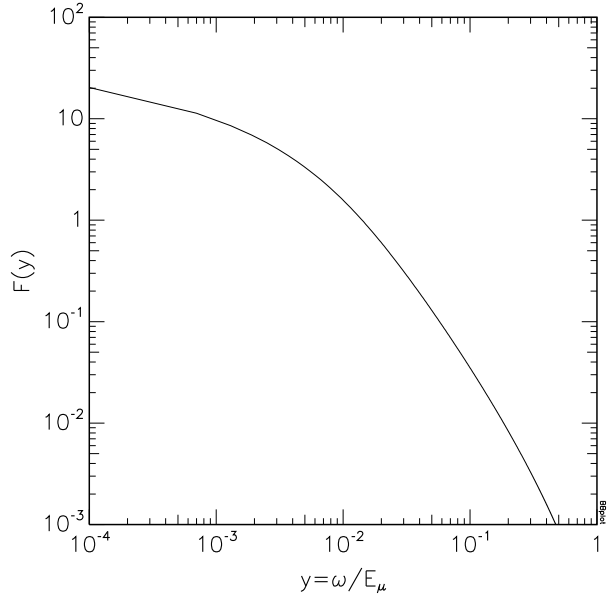,width=80mm}
\caption{Function $F(y)$ defined in Eq.~(\ref{dsigcouly}) vs.
the fractional energy loss  of the muon}
\end{center}
\end{figure}
At small $z\ll1$ (where $y\ll m_e/m_\mu$) the spectrum is logarithmically enhanced:
\begin{equation}
F(y)=\left( \frac{7}{3}+ \frac{8 z} {15} \right) 
\ln \frac{1}{z} + \frac{20}{9} 
+ \frac{511 z}{450} +\dots 
\label{asymp1}
\end{equation}
whereas at large $z \gg 1$ it is powerlike suppressed:
\begin{equation}
F(y)=\frac{1}{z} 
\left\{ (1-y) \left[ \left(2-\frac{22}{3z}\right) \ln z + 6 + \frac{5}{9 z}
\right] +
y^2 \left[ \left(2 + \frac{1}{z}\right) \ln z + 1 + \frac{1}{2 z} \right]
 \right\} + \dots
 \label{asymp2}
\end{equation}
The approximate expressions (\ref{asymp1}-\ref{asymp2}) 
agree with the exact spectrum
(\ref{dsigcouly}) within 1 \% accuracy everywhere except in the region 
$y=0.004 \div 0.02$.

From the experimental point of view of special interest is the relative mean
rate of muon energy loss due to pair production 
(or stopping power) on unit length in matter.
This quantity can be calculated as
\begin{equation}
- \frac{1}{E}\frac{dE}{dx} = n \int\limits_{2 m_e/E_\mu}^1  y \frac{d \sigma}{dy} dy =
n \;\sigma_0 \;( S_{\mathrm{Born}}+S_{\mathrm{Coul}})
\label{stop}
\end{equation}
where $n$ is the number of atoms per unit volume.
Formulae and tables for the Born contribution $S_{\mathrm{Born}}$ 
are given in Ref.~\cite{KK}.
In particular, for the case without screening
\begin{equation}
S_{\mathrm{Born}}= S_0 \left[ (1-\delta_1) \ln \frac{ E_\mu}{4 m_\mu} -1.771
\right]
\label{noscreen}
\end{equation}
and for complete screening
\begin{equation}
S_{\mathrm{Born}}= S_0 \left[ (1-\delta_1) \ln \frac{ 189}{Z^{1/3}} +0.604
\right] 
\label{screen}
\end{equation}
where
\begin{equation}
S_0=\frac{19 \pi^2}{12}\frac{m_e}{m_\mu} \,, \ \ \
\delta_1=\frac{48}{19 \pi^2} \frac{m_e}{m_\mu} 
\left( \ln\frac{m_\mu}{m_e}\right)^2=0.0352 \,.
\end{equation}
From Eqs.~(\ref{dsigcouly}) and (\ref{stop}) we derive the Coulomb correction 
\begin{equation}
S_{\mathrm{Coul}}= - f(\nu)\;\int\limits_0^1 F(y) dy = 
- \left(1-\delta_1 \right) \; f(\nu) \; S_0 \,.
\label{scoul}
\end{equation}
In the integration we have used as lower limit zero, since the region near the threshold  
$y_{\min}= 2 m_e/E_\mu$ can be safely neglected.

\section {Discussion}

In this paper we have presented the Coulomb correction to  pair
production of high energy muons on atoms. Differently to the Born contributions
of various distributions,
these corrections do not depend on the target properties, except of a simple
factor $\sigma_0 f(\nu)$ (if we do not consider the exceptional case of
pair energies close to the muon energy).

To demonstrate the relative importance of CC we discuss two simple examples:
Firstly, we present in Fig.~3 the ratio of
the spectral distribution $d \sigma_{\mathrm{Coul}}/dy$ 
to the corresponding Born cross section  
\begin{equation}
\frac{d \sigma_{\mathrm{Coul}}/dy}{d \sigma_{\mathrm{Born}}/dy}=
- \frac{ f(\nu) F(y)}{12 F_a(y,E_\mu)}
\end{equation}
where the universal function 
$F(y)$ is given in Eq.~(\ref{dsigcouly}), and values for $F_a(y,E_\mu)$ are 
taken from
Table I of \cite{KK} for collisions of 
muons with energy $E_\mu=86.4$ GeV on lead
target, $f(\nu)=0.331$. 
\begin{figure}[!htb]
\begin{center}
\epsfig{file=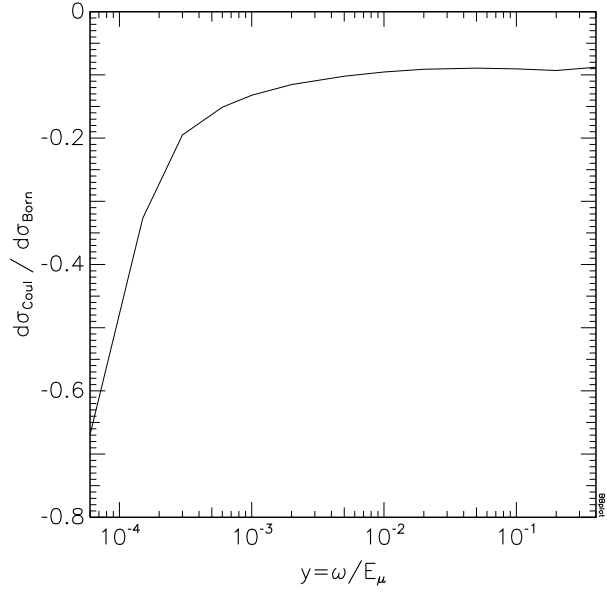,width=80mm}
\caption{Ratio of Coulomb to Born energy distribution  vs. energy fraction 
$y$ for muon collisions on Pb target at $E_\mu=86.4$ GeV}
\end{center}
\end{figure}
The presented ratio varies from about $-65$ \% to roughly $-10$ \% in the 
considered interval of pair energies.

Secondly, we compare the stopping power $S_{\mathrm{Coul}}$ with the Born term
in the two limiting cases of Eqs.~(\ref{noscreen}) and (\ref{screen}).
For muon scattering on a Pb target 
$S_{\mathrm{Coul}}/S_{\mathrm{Born}}=-15$ \% 
at $E_\mu=25 $ GeV without screening and $-7.7 $ \% for the case of complete 
screening.

It might be useful to present a simple expression for the contribution to 
$\sigma_{\mathrm{Coul}}$ above some cut $\omega>\omega_{\mathrm{cut}}$ where
this cut is restricted to the region $2 m_e \ll \omega_{\mathrm{cut}}
\ll m_e E_\mu/m_\mu$.
From Eq.~(\ref{dsigcouly}) we obtain
\begin{equation}
\sigma_{\mathrm{Coul}}(\omega_{\mathrm{cut}})=
-\frac{7}{3} \sigma_0 \; f(\nu) \left( l^2 +\frac{20}{21} l + \frac{101}{63}
\right)\,, \ \ \ l=\ln\frac{ m_e E_\mu}{m_\mu \omega_{\mathrm{cut}}} \,.
\label{sigmacut}
\end{equation}
The expression (\ref{dsigcouly}) does not remain valid close to the threshold 
$\omega_{\min}=2 m_e$. Therefore, from Eq.~(\ref{sigmacut}) 
the Coulomb correction to the total pair production cross section can be 
obtained only in leading logarithmic approximation
 choosing $\omega_{\mathrm{cut}}=2 m_e$:
\begin{equation}
\sigma_{\mathrm{Coul}}=- \frac{28}{9 \pi} \frac{ Z^2 \alpha^4}{m_e^2} \; f(\nu) 
\left( \ln \frac{ E_\mu}{ 2 m_\mu} \right)^2 \,.
\end{equation}

Finally, let us discuss the case when the muon projectile is replaced
by other charged projectiles such as electron, pion or proton.
For an electron projectile the distributions (\ref{dsigcoul}) and (\ref{dsigcouly})
remain valid changing $m_\mu \to m_e$. However, in these distributions as well
as in the Born contributions one has to take into account the effect of
the identity of the final state electrons 
and the bremsstrahlung mechanism of the $e^+e^-$--pair production
(according to \cite{BFKK} this changes the result only slightly).

For pion and proton projectiles in the basic formulae the number of photons
should be changed (besides the trivial mass replacements).
The numbers of photons are given by Eqs.~(\ref{numbers}) 
with $C= 0$, $ D= F^2_\pi(Q^2)$ for pion and $C=G^2_M(Q^2)$, 
$D= [ 4 m_p^2 G^2_E(Q^2) + Q^2 G^2_M(Q^2)]/(4 m_p^2 +Q^2)$ for proton.
Here $F_\pi$, $G_E$ and $G_M$ are the pion, proton electric and proton magnetic
form factors, respectively, $m_p$ is the proton mass.
For the pion case
these changes are essential only for $y$ close to 1 where we
should take into account the nucleus form factor, too.
For the proton case the nucleus form factor becomes important 
for somewhat smaller $y$ where the influence of the proton 
form factors is still small.

\section*{Acknowledgements}

V.G.S. is very grateful to F.~Palombo for useful discussions and warm
hospitality at INFN.
E.A.K.  is supported by INTAS grant 93-239 ext.  
V.G.S. acknowledges the fellowship by the Italian Ministry of
Foreign Affairs and support from Volkswagen Stiftung (Az. No. 1/72
302).
D.I. acknowledges support from the German BMBF under grant Nr. 05 7LP91 P0.


\begin{thebibliography}{99}

\bibitem{LL}
L.D.~Landau and E.M.~Lifshitz, Sov. Phys. {\bf 6}, 244 (1934).  

\bibitem{R} 
G.~Racah, Nuovo Cim. {\bf 14}, 93 (1937).

\bibitem{K} 
S.R.~Kel'ner, Sov. J. Nucl. Phys. {\bf 5}, 778 (1967).


\bibitem{KK}
S.R.~Kel'ner and Yu.D.~Kotov, Sov. J. Nucl. Phys. {\bf 7}, 237 (1968).  

\bibitem{N}  
A.I.~Nikishov, Sov. J. Nucl. Phys. {\bf 27}, 677 (1978).

\bibitem{PDG} 
R.M.~Barnett et al., Phys. Rev. {\bf D54}, 1 (1996) 
and 1997 off-year partial update for the 1998 edition available on 
the PDG WWW pages (URL: http://pdg.lbl.gov/).

\bibitem{G} 
A. Van Ginneken, Nucl. Instr. and Meth. {\bf A251}, 21 (1986).

\bibitem{BM}
H.A.~Bethe, L.C.~Maximon, Phys. Rev. {\bf 93}, 768 (1954);
H.~Davies, H.A.~Bethe, L.C.~ Maximon, Phys. Rev. {\bf 93}, 788 (1954).



\bibitem{AB}  
A.I.~Akhiezer, V.B.~Berestetskii: Quantum electrodynamics
(Nauka, Moscow 1981).

\bibitem{BLP}  
V.B.~Berestetskii, E.M.~Lifshitz, L.B.~Pitaevskii, Quantum
Electrodynamics (Nauka, Moscow, 1989).

\bibitem{NP}
  A.I.~Nikishov, N.V.~Pichkurov, Yad. Fiz. {\bf 35}, 964 (1982)
[Sov. J. Nucl. Phys. {\bf 35}, 561 (1982)].

\bibitem{BB}
  C.A.~Bertulani, G.~Baur, Phys. Rep. {\bf 163}, 299 (1988), Sect. 7.3.


\bibitem{IM}  
D.~Ivanov, K.~Melnikov,
Phys. Rev. {\bf D57}, 4025 (1998).

\bibitem{BGMS}  
V.M.~Budnev, I.F.~Ginzburg, G.V.~Meledin, V.G.~Serbo, Phys.
Rep. {\bf C15}, 181 (1975).

 \bibitem{BFKK}
V.N.~Baier, V.S.~Fadin, V.A.~Khoze, E.A.~Kuraev, Phys. Rep. {\bf
78}, 293(1981).

\end{thebibliography}
\end{document}